\documentclass[%
 reprint,superscriptaddress,floatfix,
 amsmath,amssymb,prl,longbibliography]{revtex4-2}

\usepackage{array}[=2023-11-01]
\usepackage{physics}
\usepackage{xcolor}
\usepackage{bm}
\usepackage{xtab,afterpage,longtable}
\usepackage{lipsum}
\usepackage{graphicx}
\usepackage{dcolumn}
\usepackage{booktabs} 
\usepackage{multirow}
\usepackage{color}
\usepackage{eucal}
\usepackage{mathrsfs}  
\usepackage{siunitx}
\usepackage{makecell}
\usepackage{hyperref}

\newcommand{\br}{\bm{r}}
\newcommand{\bq}{\bm{q}}
\newcommand{\bk}{\bm{k}}

\newcommand{\ua}{u_{\alpha}}
\newcommand{\Ua}{U_{\alpha}}

\newcommand{\E}{\mathcal{E}}
\renewcommand{\P}{\mathcal{P}}
\newcommand{\M}{\mathcal{M}}

\newcommand{\bU}{\bm{U}}
\renewcommand{\AA}{\si{\angstrom}}

\begin{document}
\raggedbottom

\title{Flexo-Photoresponses from First Principles}

\author{Zhichao Guo}
\affiliation{Department of Physics, City University of Hong Kong, Kowloon, Hong Kong SAR, China}

\author{Hua Wang}%
\email{daodaohw@zju.edu.cn}
\affiliation{Center for Quantum Matter, School of Physics, Zhejiang University, Hangzhou 310058, China}

\author{Haowei Xu}%
\email{haoweixu@cityu.edu.hk}
\affiliation{Department of Physics, City University of Hong Kong, Kowloon, Hong Kong SAR, China}

\begin{abstract}
Many functionally important materials, such as silicon, possess high-symmetry crystal structures that forbid even-order nonlinear optical responses, thereby limiting their application in e.g., photonics and optoelectronics. Strain gradients can effectively break symmetries, activating otherwise forbidden optical responses. Although such flexo-photoresponses have attracted considerable interest recently, a comprehensive microscopic theoretical framework is still lacking. Here, we develop a first-principles approach that can systematically evaluate a wide range of flexo-photoresponses in realistic materials. By mapping strain gradients to long-wavelength phonons within the established theory of electron-phonon coupling, our approach  captures the nonlocal quantum-mechanical effects of lattice deformations on nonlinear optical responses. Applying the framework to prototypical centrosymmetric materials, including Si, cubic $\rm SrTiO_3$, and monolayer black phosphorus, we demonstrate that a strain gradients of $10^5\sim10^6~\rm m^{-1}$ can induce flexo-second-harmonic generation and flexo-bulk photovoltaic effects comparable to the intrinsic responses of non-centrosymmetric crystals. Our work establishes a rigorous foundation for strain gradient engineering of light–matter interactions, opening new avenues for nanoscale optomechanics, optoelectronics, and beyond.
\end{abstract}

\maketitle

\emph{Introduction.} In addition to well-established strain engineering~\cite{zhu2010ultra,li2014elastic,shi2019deep}, \emph{strain gradient} engineering~\cite{feng2012strain,wang2020flexoelectronics,abdollahi2019converse} has attracted growing interest. This interest is driven by the increasingly large strains and strain gradients achievable in nanoscale structures and devices~\cite{dang2021achieving,banerjee2018ultralarge}, together with the ability of strain gradients to break certain crystal symmetries, most notably inversion symmetry. Such symmetry breaking can activate otherwise forbidden electric polarization (i.e., flexoelectricity~\cite{zubko2013flexoelectric}), magnetoelectric effects~\cite{liu2024flexomagnetoelectric}, as well as even-order nonlinear optical responses, hereafter termed flexo-photoresponses~\cite{yang2018flexo,zou2019boosting,wang2024flexophotovoltaic,sun2024strain,yu2024flexoelectric,jiang2021flexo}. Notably, both the direction and magnitude of these responses can be tuned on-demand through applied strain gradients. Given that many technologically important materials, such as silicon, are centrosymmetric, exploiting strain-gradient-induced flexo-photoresponses offers a powerful and versatile route to expand their functionality in optoelectronics, photonics, and related fields.

Nevertheless, the theoretical description of flexo-responses remains challenging, as strain gradients break the translational symmetry underlying conventional plane-wave  \textit{ab initio} calculations. Previously, a variety of approaches have been developed to compute flexoelectric coefficients~\cite{hong2010flexoelectricity,hong2011first,hong2013first,stengel2013flexoelectricity,maranganti2009atomistic,royo2019first,royo2022lattice}. Compared with linear flexoelectric effects, flexo-photoresponses represent higher-order nonlinear phenomena and are considerably more complicated. To date, most, if not all, first-principles studies on flexo-photoresponses utilized supercells with prescribed strain gradients~\cite{chen2025strain,xu2025competitive}. However, this supercell approach is computationally demanding and cannot accommodate general, arbitrary strain gradients. 

\begin{figure}
    \centering
    \includegraphics[width=0.6\linewidth]{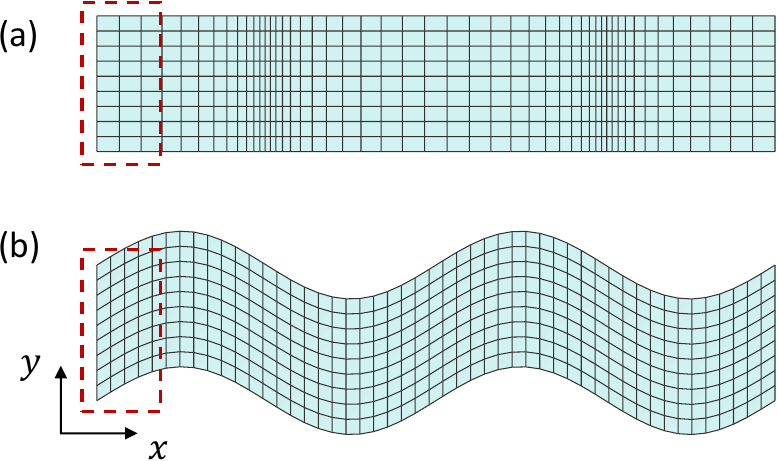}
    \caption{Illustration of the atomic displacement fields described by Eq.~(\ref{eq:atom_displacement}). The wave vector $\bq$ lies along the $x$ direction, whereas the displacement $\bU$ lies along (a) the $x$ direction and (b) the $y$ direction, corresponding to the $\eta_{xxx}$ and $\eta_{yxx}$ components of the strain gradient near $\br=0$ (red dashed box), respectively. The strain gradient can break certain crystal symmetries, particularly inversion symmetry.}
    \label{fig:displacement}
\end{figure}

In this work, we develop a systematic first-principles framework that can apply to various types of flexo-photoresponses~\cite{stengel2013flexoelectricity,baroni2001phonons}.  Starting from the unperturbed periodic crystal lattice, we represent strain gradients as long-wavelength phonon modes (Fig.~\ref{fig:displacement}), so that the influence of strain gradients can be incorporated by electron-phonon coupling. Utilizing a real-space Wannier basis, our formulation treats the electronic structure~\cite{mostofi2008wannier90}, electron-phonon coupling~\cite{giustino2007electron,lee2023electron}, and nonlinear optical responses~\cite{xu2021colossal,xu2021pure} on an equal footing. In principle, this framework universally applies to diverse photoresponses, arbitrary strain-gradient tensor components, and materials of any crystal symmetry. 

In the following, we first describe our theoretical framework for flexo-photoresponses. We then apply it to the bulk photovoltaic effect (BPVE)~\cite{dai2023recent} in diamond-cubic silicon (\textit{c}Si) and second-harmonic generation (SHG)~\cite{sipe2000second} in cubic $\rm SrTiO_3$ (\textit{c}STO), which are two prototypical centrosymmetric materials. The framework is subsequently extended to incorporate internal-strain effects (i.e., relaxation of atoms under strain gradients), and monolayer black phosphorus (\textit{b}P) is used as a testbed to compare the clamped-ion and relaxed-ion flexo-BPVE. Our calculations suggest that flexo-photoresponses in centrosymmetric materials can reach magnitudes comparable to the intrinsic photoresponses of typical non-centrosymmetric materials for strain gradients of $10^5\sim10^6~\rm m^{-1}$. This provides a quantitative design principle for strain-gradient engineering.

\begin{table*}
    \centering
  \renewcommand{\arraystretch}{1.5}
    \begin{tabular}{ccc} \hline  \hline

         &  \makecell{Flexo-coefficients of \\ centrosymmetric materials} & \makecell{Intrinsic response of \\ non-centrosymmetric materials} \\ \hline \hline
         Electric polarization&  $\sim 10^{-3}~\rm \mu C/m$~\cite{zubko2013flexoelectric,mizzi2022experimental}& $\sim 10^{3}~\rm \mu C/m^{2}$~\cite{teague1970dielectric,xu2013ferroelectric}\\ \hline  
         Bulk photovoltaic effect&  $\sim 10^{-4}~\rm m\cdot \mu A/V^2$ (this work)& $\sim 10~\rm \mu A/V^2 $~\cite{dai2023recent}\\ \hline
         Second-harmonic generation&  $\sim 10^{-5}~\rm m\cdot pm /V$ (this work)& $\sim 1~\rm pm/V$~\cite{shen1984principles,boyd2008nonlinear}\\ \hline \hline
    \end{tabular}
    \caption{Comparison between the magnitude of flexo-responses of centrosymmetric materials and intrinsic responses of non-centrosymmetric materials. Generally, a strain gradients of $10^5\sim10^6~\rm m^{-1}$ can induce flexo-responses in centrosymmetric materials comparable to the intrinsic responses of non-centrosymmetric crystals}
    \label{tab:comparison}
\end{table*}

\emph{Long-wavelength phonon formulation of flexo-photoresponses.} Unlike uniform strain, which preserves translational invariance, even a constant strain gradient breaks lattice translational symmetry and cannot be directly implemented under periodic boundary conditions. This causes significant challenges for the theoretical and computational predictions of flexo-photoresponses. To address this issue, we approximate strain gradients using long-wavelength phonon modes. Specifically, the displacement $u(\br)$ of an atom at $\br$ can be expressed as~\cite{stengel2013flexoelectricity}
\begin{equation}\label{eq:atom_displacement}
    \ua (\bm{r}) = \Ua e^{i \bq \cdot \br},
\end{equation}
where $\ua$ is the displacement along the $\alpha$ direction with amplitude $\Ua$, and $\bq$ is the phonon wave vector (Fig.~\ref{fig:displacement}). Note that here we do not consider the relaxation of atom positions under strain gradient (i.e., the clamped-ion approximation), so that all atoms within a unit cell follow the same displacement field. Internal atomic relaxation will be considered in the \textit{Relaxed-ion responses} section. From Eq.~(\ref{eq:atom_displacement}), the unsymmetrized strain $\varepsilon_{\alpha \beta}$ and strain gradient $\eta_{\alpha \beta \gamma}$ can be expressed as
\begin{equation}\label{eq:strain_strain_gradient}
\begin{aligned}
    \varepsilon_{\alpha \beta} (\br) & = \frac{\partial \ua}{\partial r_{\beta}} = i \Ua q_{\beta} e^{i \bq \cdot \br} \\
    \eta_{\alpha \beta \gamma} (\br) & = \frac{\partial^2 \ua}{\partial r_{\beta} \partial r_{\gamma}} = - \Ua q_{\beta} q_{\gamma} e^{i \bq \cdot \br} \\
\end{aligned}
\end{equation}
Here $\beta, \gamma$ are Cartesian indices. Clearly, the strain gradient is $-\Ua q_{\beta} q_{\gamma}$ for $\bq\cdot \br\ll 1$, and a uniform strain gradient corresponds to the long wavelength limit of $\bq \to 0$.

To identify the flexo-photoresponse, we need to calculate a generic third-rank tensor $\Theta^{\lambda \mu \nu}$ describing a second-order nonlinear optical response as a function of $\bU$ and $\bq$. Here, $\mu$ and $\nu$ denote the polarizations of the driving electric fields, whereas $\lambda$ labels the direction of the induced response, i.e., the photocurrent for BPVE and the second-harmonic polarization for SHG. We then expand the response around the unstrained configuration for small $\bU$ and $\bq$~\cite{hong2011first,stengel2013flexoelectricity}
\begin{equation}\label{eq:taylor_expansion}
\begin{aligned}
& \Theta^{\lambda \mu \nu}[\bq, \bU] \\
\approx  & \, \Theta^{\lambda \mu \nu,(0)} - i \sum_{\alpha, \beta} \Ua  q_{\beta} \Theta^{\lambda \mu \nu,(1)}_{\alpha \beta} - \frac{1}{2!}\sum_{\alpha, \beta, \gamma} \Ua q_{\beta}  q_{\gamma} \Theta^{\lambda \mu \nu,(2)}_{\alpha \beta \gamma}
\end{aligned}
\end{equation}
Here, $\Theta^{(i)}$ denotes the $i$th-order term in the wave-vector expansion of the response to the displacement field. Comparing Eqs.~(\ref{eq:strain_strain_gradient}) and~(\ref{eq:taylor_expansion}), one can see that $\Theta^{(0)}$ is the intrinsic photoresponse of the unstrained system, whereas $\Theta^{(1)}$ and $\Theta^{(2)}$ are the response coefficients associated with strain $\varepsilon_{\alpha \beta}$ and strain gradient $\eta_{\alpha \beta \gamma}$. They therefore describe the piezo- and flexo-photoresponses, respectively. For brevity, we will occasionally omit the superscripts $(i)$ below when the expansion order is clear from the number of subscripts.

Crystal symmetries impose selection rules on the response functions in Eq.~(\ref{eq:taylor_expansion}). Among these, inversion symmetry plays a pivotal role in governing second-order nonlinear optical effects. Under inversion operation $\P$, electric fields $\mathcal{E}$ and electric currents $j$  both acquire a minus sign. Hence, for both BPVE and SHG, one has $\P \Theta^{\lambda \mu \nu, (0)}=-\Theta^{\lambda \mu \nu, (0)}$, which enforces that the intrinsic response is $\Theta^{\lambda \mu \nu, (0)}=0$ in centrosymmetric systems. Considering strain and strain gradients, the indices $\alpha$, $\beta$, and $\gamma$ all correspond to polar vectors: $\alpha$ originates from the atomic displacement $\bU$, whereas $\beta$ and $\gamma$ originate from the phonon wave vector $\bq$. Consequently, these indices contribute additional minus signs under $\P$, yielding
\begin{equation}\label{eq:inversion_symmetry}
\begin{aligned}
  \P \Theta^{\lambda \mu \nu, (1)}_{\alpha \beta} & = -\Theta^{\lambda \mu \nu, (1)}_{\alpha \beta} \\
  \P \Theta^{\lambda \mu \nu, (2)}_{\alpha \beta \gamma} & = + \Theta^{\lambda \mu \nu, (2)}_{\alpha \beta \gamma}
\end{aligned}
\end{equation}
Therefore, in centrosymmetric systems, the piezo-photoresponses $\Theta^{\lambda \mu \nu, (1)}_{\alpha \beta}$ are forbidden, whereas the flexo-photoresponses $\Theta^{\lambda \mu \nu, (2)}_{\alpha \beta \gamma}$ can be allowed. In other words, uniform strain cannot break inversion symmetry; in contrast, a strain gradient explicitly breaks this symmetry, thereby unleashing second-order nonlinear optical responses.

In addition, other crystal symmetries like rotational symmetry and mirror symmetry further constrain the allowed tensor components. 
We will illustrate these constraints with specific examples below. In general, selection rules for $\Theta^{(i)}$ follow directly from group theory.

\emph{First-principles implementation.}
A first-principles implementation to calculate flexo-photoresponses $\Theta^{(2)}$ requires evaluating how strain gradients modify the electronic structure. This modification is naturally captured by electron-phonon coupling because the long-wavelength acoustic perturbation couples directly to the electronic states. Microscopically, strain-gradient-induced inversion-symmetry breaking is encoded in the perturbed electronic wave functions. Within nondegenerate perturbation theory, the wave-function correction due to phonon excitation can be expressed as~\cite{baroni2001phonons}
\begin{equation}\label{eq:wf_perturbation}
    \vert \Delta \phi_{n\bk} \rangle = \sum_{m} \frac{\langle \phi_{m,\bk+\bq} \vert \Delta V^{\bq}_{\rm ep} \vert \phi_{n\bk} \rangle}{\epsilon_{n\bk} - \epsilon_{m,\bk+\bq} + i\zeta} \vert \phi_{m,\bk+\bq} \rangle
\end{equation}
Here, $m$ and $n$ are band indices, $\vert\phi_{m\bk}\rangle$ is an eigenstate of the unperturbed Hamiltonian in band $m$ at wave vector $\bk$, and $\epsilon_{m\bk}$ is the corresponding band energy. The state $\vert\Delta\phi_{n\bk}\rangle$ denotes the first-order correction to the wave function, where the change in the self-consistent potential, $\Delta V^{\bq}_{\rm ep}$, arises from electron-phonon coupling when phonons are excited. Crucially, the strain-gradient-induced correction $\vert \Delta \phi_{n\bk} \rangle$ can directly break the inversion-symmetry relations among electronic states, thereby activating otherwise forbidden second-order photoresponses. This correction is evaluated within a unified Wannier representation of both the electronic states $\vert\phi_{n\bk}\rangle$ and electron-phonon coupling matrix elements $\Delta V^{\bq}_{\rm ep}$. The perturbed states, $\vert\phi_{n\bk}\rangle + \vert \Delta \phi_{n\bk} \rangle$, are subsequently used to compute the optical matrix elements and photoresponses, following standard formulations of nonlinear optical responses. Detailed derivations are provided in the Supplemental Material.

Note that there can be (near) degeneracy in band energies, i.e., $\epsilon_{n\bk} \approx \epsilon_{m,\bk+\bq}$, especially when $\bq$ is small. In this case, degenerate perturbation theory is generally required. In the present implementation, we introduce a phenomenological electron linewidth $\zeta=20~\rm meV$ to regularize the numerical divergence when $\epsilon_{n\bk}\approx\epsilon_{m,\bk+\bq}$. Our numerical results are relatively insensitive to $\zeta$. 

In the following sections, we apply this approach to evaluate flexo-BPVE and flexo-SHG in representative material systems.

\begin{figure}
    \centering
    \includegraphics[width=1\linewidth]{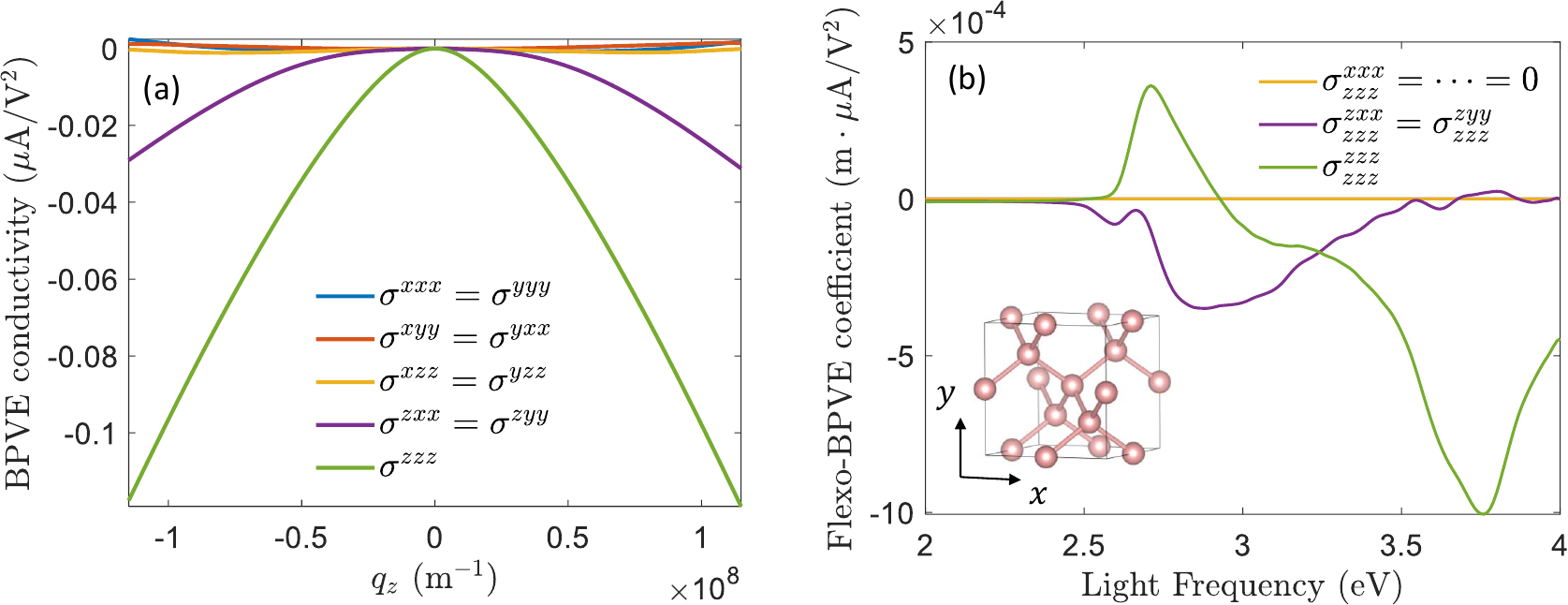}
    \caption{(a) BPVE conductivity $\sigma^{\lambda \mu \nu}[\omega; \boldsymbol{q}, \boldsymbol{U}]$ as a function of $q_z$ for \textit{c}Si, whose atomic structure is shown in the inset of (b). Here we fix $\omega = 3.5~\rm eV$ and $U_x = U_y = 0, \, U_z = 5\times10^{-4}~\rm \AA$. Due to the equivalence of $x$ and $y$, $\sigma^{\lambda \mu \nu}$ is invariant under the exchange of $x\leftrightarrow y$. (b) Flexo-BPVE coefficients of \textit{c}Si as a function of light frequency $\omega$. }
    \label{fig:silicon}
\end{figure}

\emph{Flexo-bulk photovoltaic effect.} We first consider the BPVE, a second-order nonlinear optical response described by
\begin{equation}\label{eq:bpv_definition}
\begin{aligned}
  j^{\lambda} = \sigma^{\lambda \mu \nu}[\omega; \boldsymbol{q}, \boldsymbol{U}] \E^{\mu}(\omega) \E^{\nu}(-\omega) 
\end{aligned}
\end{equation}
Specifically, the $+\omega$ and $-\omega$ frequency components of the incident light field $\E$ generate a spatially uniform direct charge current $j^{\lambda}$. The formula for calculating the conductivity $\sigma^{\lambda \mu \nu}[\omega;\boldsymbol{q},\boldsymbol{U}]$ is provided in the End Matter.

We examine the BPVE in diamond-cubic Si (\textit{c}Si), a prototypical centrosymmetric semiconductor. We consider $\alpha=\beta=\gamma=z$, with $z$ along the $[001]$ crystal direction, corresponding to the $\eta_{zzz}$ component of the strain-gradient tensor. Other components can be treated similarly. We fix the displacement along $z$ at $U_z=5\times10^{-4}~\rm \AA$ and omit the $\bU$ argument hereafter.

Figure~\ref{fig:silicon}(a) shows the photoconductivity as a function of $q_z$. In pristine \textit{c}Si, the $x$, $y$, and $z$ directions are equivalent. The strain gradient $\eta_{zzz}$ singles out the $z$ direction while preserving the equivalence of $x$ and $y$. Hence, the BPVE conductivity must be invariant under the exchange $x\leftrightarrow y$, as confirmed by the calculated spectra. A small residual value remains at $q_z=0$ because of numerical noise. We remove this constant offset from each curve before fitting, thereby setting $\sigma^{\lambda\mu\nu}[q_z=0]=0$. Near $q_z=0$, the slope $\partial\sigma^{\lambda\mu\nu}/\partial q_z$ vanishes because the first-order term $\sigma^{\lambda\mu\nu}_{zz}$ (piezo-BPVE) is forbidden by the inversion symmetry, as we described above. Away from the long-wavelength limit, $\sigma^{\lambda\mu\nu}$ is not symmetric with respect to $\pm q_z$. Here, $+q_z$ and $-q_z$ correspond to compressive and tensile strain [cf. Eq.~(\ref{eq:strain_strain_gradient})] and thus are not fully equivalent when $\vert q_z\vert$ is large.

We obtain the flexo-BPVE coefficient tensor from the second-order derivative 
\begin{equation}\label{eq:bpv_flexo}
\begin{aligned}
 \sigma^{\lambda\mu\nu}_{zzz} = -\frac{1}{U_z}\left.\frac{\partial^2\sigma^{\lambda\mu\nu}}{\partial q_z^2}\right\vert_{q_z=0}.
\end{aligned}
\end{equation}
The resulting spectra are shown in Fig.~\ref{fig:silicon}(b) as a function of light frequency $\omega$. Three components $\sigma^{zzz}_{zzz}, $ $\sigma^{zxx}_{zzz}$, and $\sigma^{zyy}_{zzz}$ are nonzero, with the latter two equal because $x$ and $y$ remain equivalent. Some other components, such as $\sigma^{xxx}_{zzz}$, vanish throughout the plotted frequency range. They are forbidden by some crystal symmetries that survive under the strain gradient $\eta_{zzz}$. Particularly, pristine \textit{c}Si has the combined mirror symmetries $\M_x\M_y$, $\M_x\M_z$, and $\M_y\M_z$. Here, $\M^{\xi}$ with $\xi$  a Cartesian index indicates mirror operation with respect to a plane perpendicular to the $\xi$ direction, i.e., $\xi \to -\xi$. The strain gradient $\eta_{zzz}$ breaks the combinations involving $\M_z$ but preserves $\M_x\M_y$, which would impose selection rules on the response tensor $\sigma^{\lambda\mu\nu}_{zzz}$. For example, one has $\M_x\M_y \sigma^{xxx}_{zzz} \to -\sigma^{xxx}_{zzz}$, and therefore $\sigma^{xxx}_{zzz}=0$. 

The magnitudes of the flexo-photoreponses are crucial for their practical applications. The allowed flexo-BPVE coefficients of $c\rm Si$ reach $10^{-4}\sim10^{-3}~\rm m\cdot\mu A/V^2$. For comparison, the BPVE conductivity of typical non-centrosymmetric materials is of order $10\sim100~\rm \mu A/V^2$~\cite{young2012first,dai2023recent}. The calculated coefficients therefore indicate that a strain gradient of order $10^5~\rm m^{-1}$ is sufficient for the flexo-BPVE conductivity of \textit{c}Si to reach the characteristic magnitude of intrinsic BPVE in noncentrosymmetric materials.

\begin{figure}
    \centering
    \includegraphics[width=1\linewidth]{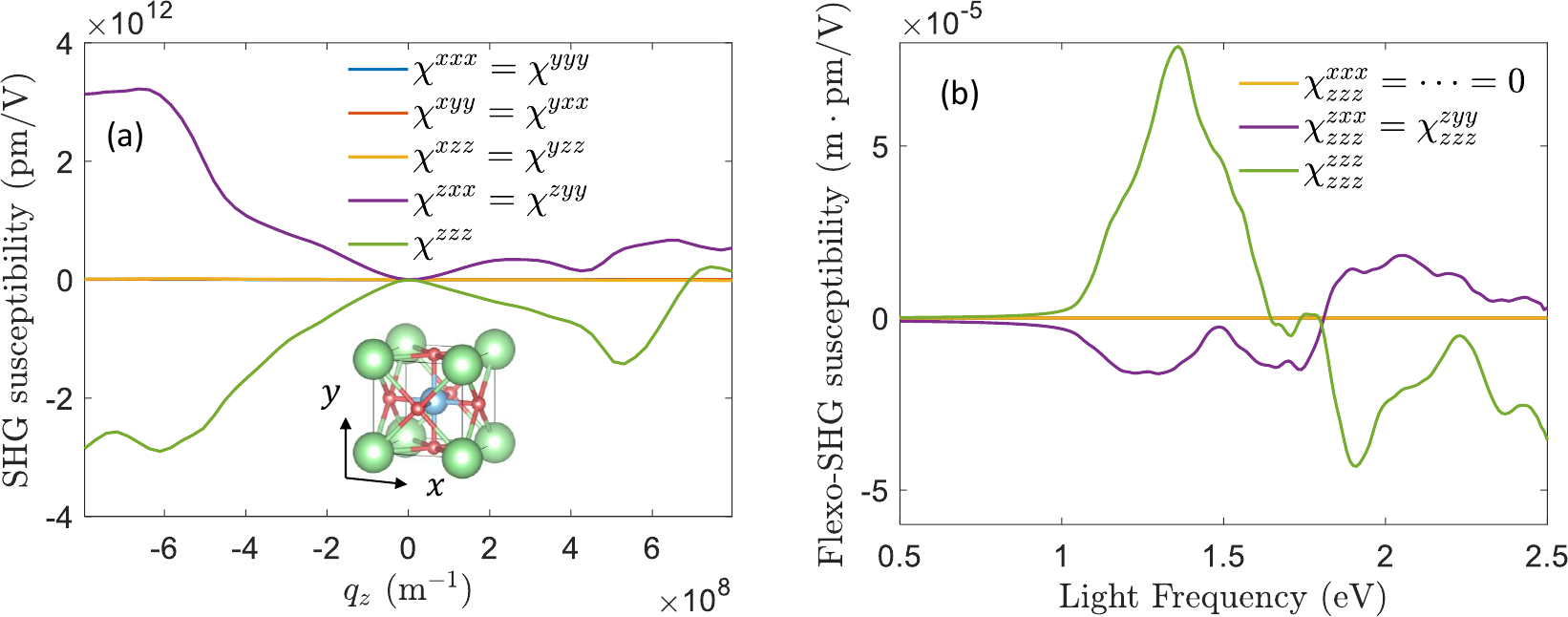}
    \caption{(a) SHG susceptibility $\chi^{\lambda \mu \nu} 
    [\omega; \boldsymbol{q}, \boldsymbol{U}]$ as a function of $q_z$ for \textit{c}STO, whose atomic structure is shown in the inset of (b). Here we fix $\omega = 2~\rm eV$ and $U_x = U_y = 0, \, U_z = 5\times10^{-4}~\rm \AA$. $\chi^{\lambda \mu \nu}$ is invariant under the exchange of $x\leftrightarrow y$. $\chi^{\lambda \mu \nu} 
    [\omega; \boldsymbol{q}, \boldsymbol{U}]$ has a quadratic dependence on $q_z$ as $q_z\to 0$. (b) Flexo-SHG susceptibility of \textit{c}STO as a function of light frequency $\omega$. }
    \label{fig:sto}
\end{figure} 

\emph{Flexo-second-harmonic generation.} Next, we examine another prototypical nonlinear optical effect, second-harmonic generation, which can be expressed as
\begin{equation}\label{eq:shg_definition}
\begin{aligned}
  E^{\lambda}(2\omega) = \chi^{\lambda \mu \nu}(\omega; \boldsymbol{q}, \boldsymbol {U}) \E^{\mu}(\omega) \E^{\nu}(\omega) \\
\end{aligned}
\end{equation}
SHG is a frequency-doubling process in which two photons of frequency $\omega$ combine to generate a third photon of frequency $2\omega$. The explicit expression used to calculate $\chi^{\lambda\mu\nu}$ is given in the End Matter.

We use cubic $\rm SrTiO_3$ (\textit{c}STO) to evaluate the flexo-SHG response. Like \textit{c}Si, pristine \textit{c}STO is centrosymmetric and has equivalent $x$, $y$, and $z$ axes. Under a strain gradient $\eta_{zzz}$, the $x$ and $y$ directions remain equivalent, as shown in Fig.~\ref{fig:sto}(a), where $\chi^{\lambda\mu\nu}$ is plotted as a function of $q_z$. The flexo-SHG susceptibility is obtained from $\chi^{\lambda\mu\nu}_{zzz}=  -\frac{1}{U_z} \left.\frac{\partial^2\chi^{\lambda\mu\nu}}{\partial q_z^2}\right\vert_{q_z=0}$ and is shown in Fig.~\ref{fig:sto}(b). Components with an odd number of $x$ or $y$ indices, such as $\chi^{xxx}_{zzz}$, vanish because they are forbidden by the mirror symmetries $\M_x$ and $\M_y$, which persist under the $\eta_{zzz}$ strain gradient. The allowed components of the flexo-SHG susceptibility are of order $10^{-5}~\rm m\cdot pm/V$. Compared with SHG susceptibilities characteristic of noncentrosymmetric materials~\cite{shen1984principles,boyd2008nonlinear}, this coefficient implies that a strain gradient of approximately $10^5\sim10^6~\rm m^{-1}$ is sufficient to produce an SHG response of comparable magnitude. This corresponds, for example, to a $1\%$ strain variation over $10\sim100~\rm nm$.

\emph{Relaxed-ion responses.} The preceding calculations are based on the clamped-ion approximation, in which atoms within a unit cell follow a common displacement field described by Eq.~(\ref{eq:atom_displacement}).  In real materials, however, different atoms may undergo additional relative displacements in response to a strain gradient, giving rise to internal strain~\cite{stengel2013flexoelectricity}. To incorporate these material-specific relaxation effects, we extend the above formalism to a relaxed-ion formulation through a site-specific Taylor expansion in the phonon wave vector $\bq$,
\begin{equation}\label{eq:relaxation}
\begin{aligned}
  U^{\kappa}_{\beta} \approx 
  \sum_{\alpha} U_{\alpha} \left[ \delta_{\beta\alpha} + \sum_{\gamma} i q_{\gamma} \Gamma^{\kappa}_{\beta \alpha \gamma} - \sum_{\gamma, \lambda} q_{\gamma} q_{\lambda} N^{\kappa}_{\beta \alpha \gamma \lambda} \right]
\end{aligned}
\end{equation}
Here, $U^{\kappa}_{\beta}$ is the relaxed displacement of atom $\kappa$ in the unit cell along the $\beta$ direction. The tensors $\Gamma^{\kappa}_{\beta\alpha\gamma}$ and $N^{\kappa}_{\beta\alpha\gamma\lambda}$ are the first- and second-order internal-strain tensors, respectively, which can be obtained from the real-space force-constant matrix (see the Supplemental Material and Ref.~\cite{stengel2013flexoelectricity}).

\begin{figure}[!t]
    \centering
    \includegraphics[width=0.5\linewidth]{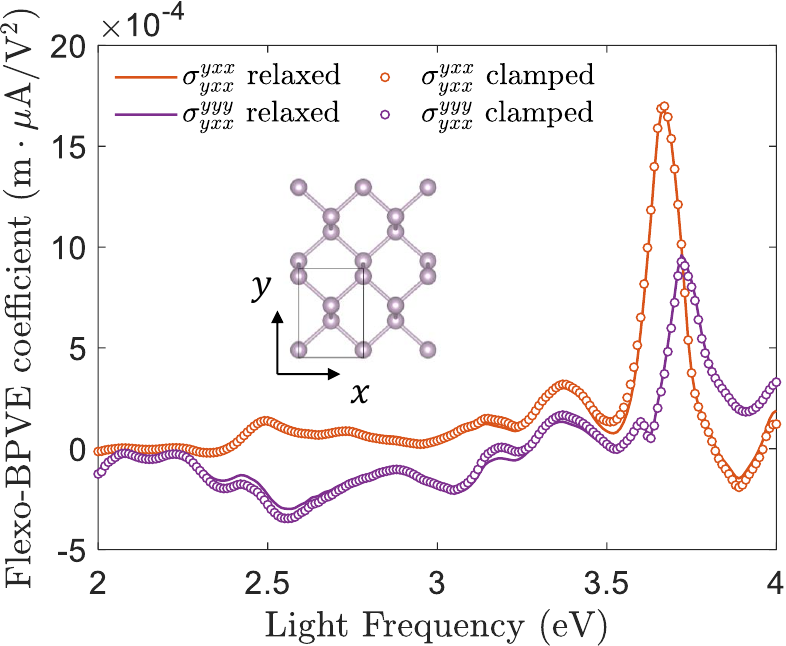}
    \caption{Flexo-BPVE coefficients of monolayer black phosphorus (atomic structure in the inset). Solid lines and circles correspond to relaxed- and clamped-ion responses, respectively.}
    \label{fig:bp}
\end{figure}

We select monolayer black phosphorus (\textit{b}P) to illustrate the influence of internal strain and the application of the relaxed-ion formulation. The nonplanar, puckered unit cell of \textit{b}P contains four phosphorus atoms and has a hinge-like structure~\cite{rodin2014strain,jiang2014negative} (inset of Fig.~\ref{fig:bp}). This geometry accommodates deformation through changes in bond and dihedral angles rather than solely through bond stretching~\cite{jiang2014negative,wei2014superior}, providing substantial internal degrees of freedom. Hence, \textit{b}P is a suitable example for comparing clamped- and relaxed-ion flexo-photoresponses.

We investigate the flexo-BPVE coefficients $\sigma^{\lambda \mu \nu}_{\alpha \beta \gamma}$ under a strain gradient $\eta_{yxx}$. Pristine \textit{b}P possesses both inversion $\P$ and mirror $\M_x$ symmetry (inset of Fig.~\ref{fig:bp}), which forbid tensor components with an odd total number of $x$ indices. Fig.~\ref{fig:bp} shows two nonzero components, $\sigma^{yxx}_{yxx}$ and $\sigma^{yyy}_{yxx}$, calculated within both the clamped- and relaxed-ion approximations, using the displacements in Eqs.~(\ref{eq:atom_displacement}) and (\ref{eq:relaxation}), respectively. For the strain-gradient component and frequency range considered here, internal strain changes the flexo-BPVE response by only about $1~\%$. That is, the clamped- and relaxed-ion calculations yield similar results for \textit{b}P. More generally, our relaxed-ion formulation provides a systematic scheme to incorporate internal-strain effects when they are significant, such as in materials with soft phonon modes or near structural phase transitions.  

\emph{Discussion and conclusions.} Our first-principles framework quantitatively evaluates the magnitude of flexo-photoresponses. For both BPVE and SHG, we determine the critical strain gradients $\eta_{\rm cri}$ required for flexo-photoresponses in centrosymmetric materials to rival intrinsic second-order optical responses in non-centrosymmetric materials. Across distinct materials and optical processes, our calculations reveal $\eta_{\rm cri}$ in the range of $10^5 \sim 10^6~\rm m^{-1}$ (Table~\ref{tab:comparison}), corresponding to a strain variation of order $1\%$ across a length scale of $10 \sim 100~\rm nm$. At this scale, the flexo-BPVE conductivity of \textit{c}Si and the flexo-SHG susceptibility of \textit{c}STO reach magnitudes comparable to those of typical non-centrosymmetric crystals. This offers quantitative design principles for the strain-gradient engineering of nonlinear optical responses. 

This scale closely parallels the flexoelectric effect: given typical flexoelectric coefficients of $1\sim10~\rm nC/m$~\cite{zubko2013flexoelectric,mizzi2022experimental}, strain gradients in the range of $10^5\sim10^6~\rm m^{-1}$ induce an electric polarization of order $1~\rm \mu C/cm^2$ in centrosymmetric materials, typical of noncentrosymmetric polar materials~\cite{teague1970dielectric,xu2013ferroelectric}. This direct analogy between flexoelectric polarization and flexo-photoresponses stems from their shared physical origin: the breaking of inversion symmetry in electronic structure due to strain gradients.

The applicability of strain-gradient engineering for nonlinear optical phenomena relies on generating steep strain gradients. Notably, increasingly large strain gradients have been realized experimentally. For example, strain gradients on the order of $10^6~\rm m^{-1}$ have been realized in a MoS$_2$ sheet~\cite{jiang2021flexo}, precisely the regime where flexo-BPVE matches the intrinsic BPVE of non-centrosymmetric crystals. 
Comparable gradients can also be generated via atomic force microscope (AFM) nano-indentation~\cite{park2020colossal,park2021flexoelectric}, dislocation engineering~\cite{gao2018atomic}, or nano-structural features such as edges and wrinkles~\cite{sun2024strain}.  These diverse, experimentally viable platforms underscore the broad feasibility of leveraging strain gradients to drive nonlinear optical responses in centrosymmetric systems, thereby unleashing their potential in e.g., photonics and optoelectronics.

In summary, we have developed a unified theoretical and computational formalism to predict flexo-photoresponses from first principles. By mapping strain gradients to long-wavelength phonon modes, their influence on electronic structures and nonlinear optical responses is naturally incorporated via electron-phonon coupling. The framework accommodates both a computationally efficient clamped-ion treatment, ideal for high-throughput materials screening or machine-learning integration, and a more rigorous relaxed-ion extension that explicitly accounts for internal atomic relaxation. In principle, our formalism applies to diverse flexo-photoresponses, arbitrary strain gradients, and materials of any crystal symmetry. Beyond the polar-vector responses like BPVE and SHG considered here, extending the computational framework, as well as symmetry analysis, to axial-vector responses, such as the nonlinear Edelstein effect~\cite{xu2021light}, represents an exciting avenue for future work. Ultimately, this approach establishes a quantitative foundation for unlocking nontrivial nonlinear optical responses in centrosymmetric materials via on-demand strain gradients.

\emph{Acknowledgements} 
Z.G. and H.X. acknowledge the support from the Hong Kong Research Grants Council under ECS Grant No. 21303526, National Natural Science Foundation of China under Grant No. 12604782, and the startup funding from City University of Hong Kong. H.W. acknowledges the support from the NSFC under Grants Nos. 12474240, 12522411, and 12304049, as well as the support from Fundamental Research Funds for the Central Universities.

\bibliography{bibliography}

@article{liu2024flexomagnetoelectric,
  title={Flexomagnetoelectric Effect in Sr 2 IrO 4 Thin Films},
  author={Liu, Xin and Hu, Ting and Zhang, Yujun and Xu, Xueli and Lei, Runyu and Wu, Biao and Ma, Zongwei and Lv, Peng and Zhang, Yuelin and Huang, Shih-Wen and others},
  journal={Physical Review Letters},
  volume={133},
  number={15},
  pages={156505},
  year={2024},
  publisher={APS}
}

@article{yu2024flexoelectric,
  title={Flexoelectric Engineering of Bulk Photovoltaic Photodetector},
  author={Yu, Junxi and Huang, Boyuan and Yang, Songjie and Zhang, Yuan and Bai, Yinxin and Song, Chunlin and Ming, Wenjie and Liu, Wenyuan and Wang, Junling and Li, Changjian and others},
  journal={Nano Letters},
  year={2024},
  publisher={ACS Publications}
}

@article{sun2024strain,
  title={Strain-Prompted Giant Flexo-Photovoltaic Effect in Two-Dimensional Violet Phosphorene Nanosheets},
  author={Sun, Ruo-Xuan and Hu, Zhen and Zhao, Xuewen and Zha, Ming-Jie and Zhang, Jinying and Chen, Xu-Dong and Liu, Zhibo and Tian, Jianguo},
  journal={ACS nano},
  year={2024},
  publisher={ACS Publications}
}

@article{xu2021light,
  title={Light-induced static magnetization: Nonlinear Edelstein effect},
  author={Xu, Haowei and Zhou, Jian and Wang, Hua and Li, Ju},
  journal={Physical Review B},
  volume={103},
  number={20},
  pages={205417},
  year={2021},
  publisher={APS}
}

@article{xu2021pure,
  title={Pure spin photocurrent in non-centrosymmetric crystals: bulk spin photovoltaic effect},
  author={Xu, Haowei and Wang, Hua and Zhou, Jian and Li, Ju},
  journal={Nature Communications},
  volume={12},
  number={1},
  pages={4330},
  year={2021},
  publisher={Nature Publishing Group UK London}
}

@article{stengel2013flexoelectricity,
  title={Flexoelectricity from density-functional perturbation theory},
  author={Stengel, Massimiliano},
  journal={Physical Review B},
  volume={88},
  number={17},
  pages={174106},
  year={2013},
  publisher={APS}
}

@article{zhu2010ultra,
  title={Ultra-strength materials},
  author={Zhu, Ting and Li, Ju},
  journal={Progress in Materials Science},
  volume={55},
  number={7},
  pages={710--757},
  year={2010},
  publisher={Elsevier}
}

@article{feng2012strain,
  title={Strain-engineered artificial atom as a broad-spectrum solar energy funnel},
  author={Feng, Ji and Qian, Xiaofeng and Huang, Cheng-Wei and Li, Ju},
  journal={Nature Photonics},
  volume={6},
  number={12},
  pages={866--872},
  year={2012},
  publisher={Nature Publishing Group UK London}
}

@article{shi2019deep,
  title={Deep elastic strain engineering of bandgap through machine learning},
  author={Shi, Zhe and Tsymbalov, Evgenii and Dao, Ming and Suresh, Subra and Shapeev, Alexander and Li, Ju},
  journal={Proceedings of the National Academy of Sciences},
  volume={116},
  number={10},
  pages={4117--4122},
  year={2019},
  publisher={National Acad Sciences}
}

@article{dang2021achieving,
  title={Achieving large uniform tensile elasticity in microfabricated diamond},
  author={Dang, Chaoqun and Chou, Jyh-Pin and Dai, Bing and Chou, Chang-Ti and Yang, Yang and Fan, Rong and Lin, Weitong and Meng, Fanling and Hu, Alice and Zhu, Jiaqi and others},
  journal={Science},
  volume={371},
  number={6524},
  pages={76--78},
  year={2021},
  publisher={American Association for the Advancement of Science}
}

@article{li2014elastic,
  title={Elastic strain engineering for unprecedented materials properties},
  author={Li, Ju and Shan, Zhiwei and Ma, Evan},
  journal={Mrs Bulletin},
  volume={39},
  number={2},
  pages={108--114},
  year={2014},
  publisher={Cambridge University Press}
}

@article{wang2020flexoelectronics,
  title={Flexoelectronics of centrosymmetric semiconductors},
  author={Wang, Longfei and Liu, Shuhai and Feng, Xiaolong and Zhang, Chunli and Zhu, Laipan and Zhai, Junyi and Qin, Yong and Wang, Zhong Lin},
  journal={Nature Nanotechnology},
  volume={15},
  number={8},
  pages={661--667},
  year={2020},
  publisher={Nature Publishing Group UK London}
}

@article{zubko2013flexoelectric,
  title={Flexoelectric effect in solids},
  author={Zubko, Pavlo and Catalan, Gustau and Tagantsev, Alexander K},
  journal={Annual Review of Materials Research},
  volume={43},
  pages={387--421},
  year={2013},
  publisher={Annual Reviews}
}

@article{banerjee2018ultralarge,
  title={Ultralarge elastic deformation of nanoscale diamond},
  author={Banerjee, Amit and Bernoulli, Daniel and Zhang, Hongti and Yuen, Muk-Fung and Liu, Jiabin and Dong, Jichen and Ding, Feng and Lu, Jian and Dao, Ming and Zhang, Wenjun and others},
  journal={Science},
  volume={360},
  number={6386},
  pages={300--302},
  year={2018},
  publisher={American Association for the Advancement of Science}
}

@article{yang2018flexo,
  title={Flexo-photovoltaic effect},
  author={Yang, Ming-Min and Kim, Dong Jik and Alexe, Marin},
  journal={Science},
  volume={360},
  number={6391},
  pages={904--907},
  year={2018},
  publisher={American Association for the Advancement of Science}
}

@article{zou2019boosting,
  title={Boosting the solar cell efficiency by flexo-photovoltaic effect?},
  author={Zou, Haiyang and Zhang, Chunli and Xue, Hao and Wu, Zhiyi and Wang, Zhong Lin},
  journal={ACS nano},
  volume={13},
  number={11},
  pages={12259--12267},
  year={2019},
  publisher={ACS Publications}
}

@article{wang2024flexophotovoltaic,
  title={Flexophotovoltaic effect and above-band-gap photovoltage induced by strain gradients in halide perovskites},
  author={Wang, Zhiguo and Shu, Shengwen and Wei, Xiaoyong and Liang, Renhong and Ke, Shanming and Shu, Longlong and Catalan, Gustau},
  journal={Physical Review Letters},
  volume={132},
  number={8},
  pages={086902},
  year={2024},
  publisher={APS}
}

@article{chen2025strain,
  title={Strain gradient engineered flexophotovoltaics and spin polarization in bent {MoSi$_2$N$_4$} nanoribbons},
  author={Chen, Helong and Ge, Meng and Xiao, Yang and Xu, Degao and Tan, Jianing and Ouyang, Gang},
  journal={Physical Review B},
  volume={112},
  number={24},
  pages={245407},
  year={2025},
  doi={10.1103/z4m4-z59l},
  publisher={APS}
}

@article{xu2025competitive,
  title={Competitive polarization-enhanced flexophotovoltaic effect in a piezoelectric {Janus In$_2$STe} monolayer},
  author={Xu, Degao and Xu, Jiahua and Chen, Helong and Ge, Meng and Dong, Jiansheng and Chen, Shanjun and Dai, Chenjie and Wang, Xianzhuo and Tan, Jianing and Yang, Wenxing and Ouyang, Gang},
  journal={Physical Review B},
  volume={112},
  number={24},
  pages={245418},
  year={2025},
  doi={10.1103/dfqf-4vxv},
  publisher={APS}
}

@article{abdollahi2019converse,
  title={Converse flexoelectricity yields large piezoresponse force microscopy signals in non-piezoelectric materials},
  author={Abdollahi, Amir and Domingo, Neus and Arias, Irene and Catalan, Gustau},
  journal={Nature communications},
  volume={10},
  number={1},
  pages={1266},
  year={2019},
  publisher={Nature Publishing Group UK London}
}

@article{hong2010flexoelectricity,
  title={The flexoelectricity of barium and strontium titanates from first principles},
  author={Hong, Jiawang and Catalan, G and Scott, JF and Artacho, E},
  journal={Journal of Physics: Condensed Matter},
  volume={22},
  number={11},
  pages={112201},
  year={2010},
  publisher={IOP Publishing}
}

@article{hong2013first,
  title={First-principles theory and calculation of flexoelectricity},
  author={Hong, Jiawang and Vanderbilt, David},
  journal={Physical Review B},
  volume={88},
  number={17},
  pages={174107},
  year={2013},
  publisher={APS}
}

@article{maranganti2009atomistic,
  title={Atomistic determination of flexoelectric properties of crystalline dielectrics},
  author={Maranganti, R and Sharma, P},
  journal={Physical Review B},
  volume={80},
  number={5},
  pages={054109},
  year={2009},
  publisher={APS}
}

@article{royo2019first,
  title={First-principles theory of spatial dispersion: Dynamical quadrupoles and flexoelectricity},
  author={Royo, Miquel and Stengel, Massimiliano},
  journal={Physical Review X},
  volume={9},
  number={2},
  pages={021050},
  year={2019},
  publisher={APS}
}

@article{royo2022lattice,
  title={Lattice-mediated bulk flexoelectricity from first principles},
  author={Royo, Miquel and Stengel, Massimiliano},
  journal={Physical Review B},
  volume={105},
  number={6},
  pages={064101},
  year={2022},
  publisher={APS}
}

@article{mostofi2008wannier90,
  title={wannier90: A tool for obtaining maximally-localised Wannier functions},
  author={Mostofi, Arash A and Yates, Jonathan R and Lee, Young-Su and Souza, Ivo and Vanderbilt, David and Marzari, Nicola},
  journal={Computer physics communications},
  volume={178},
  number={9},
  pages={685--699},
  year={2008},
  publisher={Elsevier}
}

@article{giustino2007electron,
  title={Electron-phonon interaction using Wannier functions},
  author={Giustino, Feliciano and Cohen, Marvin L and Louie, Steven G},
  journal={Physical Review B},
  volume={76},
  number={16},
  pages={165108},
  year={2007},
  publisher={APS}
}

@article{lee2023electron,
  title={Electron--phonon physics from first principles using the EPW code},
  author={Lee, Hyungjun and Ponc{\'e}, Samuel and Bushick, Kyle and Hajinazar, Samad and Lafuente-Bartolome, Jon and Leveillee, Joshua and Lian, Chao and Lihm, Jae-Mo and Macheda, Francesco and Mori, Hitoshi and others},
  journal={npj Computational Materials},
  volume={9},
  number={1},
  pages={156},
  year={2023},
  publisher={Nature Publishing Group UK London}
}

@article{xu2021colossal,
  title={Colossal switchable photocurrents in topological Janus transition metal dichalcogenides},
  author={Xu, Haowei and Wang, Hua and Zhou, Jian and Guo, Yunfan and Kong, Jing and Li, Ju},
  journal={npj Computational Materials},
  volume={7},
  number={1},
  pages={31},
  year={2021},
  publisher={Nature Publishing Group UK London}
}

@article{dai2023recent,
  title={Recent progress in the theory of bulk photovoltaic effect},
  author={Dai, Zhenbang and Rappe, Andrew M},
  journal={Chemical Physics Reviews},
  volume={4},
  number={1},
  year={2023},
  publisher={AIP Publishing}
}

@article{sipe2000second,
  title={Second-order optical response in semiconductors},
  author={Sipe, JE and Shkrebtii, AI},
  journal={Physical Review B},
  volume={61},
  number={8},
  pages={5337},
  year={2000},
  publisher={APS}
}

@article{baroni2001phonons,
  title={Phonons and related crystal properties from density-functional perturbation theory},
  author={Baroni, Stefano and De Gironcoli, Stefano and Dal Corso, Andrea and Giannozzi, Paolo},
  journal={Reviews of modern Physics},
  volume={73},
  number={2},
  pages={515},
  year={2001},
  publisher={APS}
}

@article{hong2011first,
  title={First-principles theory of frozen-ion flexoelectricity},
  author={Hong, Jiawang and Vanderbilt, David},
  journal={Physical Review B},
  volume={84},
  number={18},
  pages={180101},
  year={2011},
  publisher={APS}
}

@article{mizzi2022experimental,
  title={Experimental determination of flexoelectric coefficients in SrTiO 3, KTaO 3, TiO 2, and YAlO 3 single crystals},
  author={Mizzi, Christopher A and Guo, Binghao and Marks, Laurence D},
  journal={Physical Review Materials},
  volume={6},
  number={5},
  pages={055005},
  year={2022},
  publisher={APS}
}

@book{shen1984principles,
  title={Principles of nonlinear optics},
  author={Shen, Yuen-Ron},
  year={1984},
  publisher={Wiley-Interscience, New York, NY, USA}
}

@incollection{boyd2008nonlinear,
  title={Nonlinear optics},
  author={Boyd, Robert W and Gaeta, Alexander L and Giese, Enno},
  booktitle={Springer Handbook of Atomic, Molecular, and Optical Physics},
  pages={1097--1110},
  year={2008},
  publisher={Springer}
}

@book{xu2013ferroelectric,
  title={Ferroelectric materials and their applications},
  author={Xu, Yuhuan},
  year={2013},
  publisher={Elsevier}
}

@article{teague1970dielectric,
  title={Dielectric hysteresis in single crystal BiFeO3},
  author={Teague, James R and Gerson, Robert and James, William Joseph},
  journal={Solid State Communications},
  volume={8},
  number={13},
  pages={1073--1074},
  year={1970},
  publisher={Elsevier}
}

@article{sharma2003linear,
  title={Linear and second-order optical response of III-V monolayer superlattices},
  author={Sharma, S and Dewhurst, J Kay and Ambrosch-Draxl, C},
  journal={Physical Review B},
  volume={67},
  number={16},
  pages={165332},
  year={2003},
  publisher={APS}
}

@article{jiang2014negative,
  title={Negative {Poisson}'s ratio in single-layer black phosphorus},
  author={Jiang, Jin-Wu and Park, Harold S.},
  journal={Nature Communications},
  volume={5},
  pages={4727},
  year={2014},
  doi={10.1038/ncomms5727}
}

@article{jiang2021flexo,
  title={Flexo-photovoltaic effect in {MoS}$_2$},
  author={Jiang, Jie and Chen, Zhizhong and Hu, Yang and Xiang, Yu and Zhang, Lifu and Wang, Yiping and Wang, Gwo-Ching and Shi, Jian},
  journal={Nature Nanotechnology},
  volume={16},
  number={8},
  pages={894--901},
  year={2021},
  publisher={Nature Publishing Group}
}

@article{park2020colossal,
  title={Colossal flexoresistance in dielectrics},
  author={Park, Sung Min and Wang, Bo and Paudel, Tula and Park, Se Young and Das, Saikat and Kim, Jeong Rae and Ko, Eun Kyo and Lee, Han Gyeol and Park, Nahee and Tao, Lingling and Suh, Dongseok and Tsymbal, Evgeny Y. and Chen, Long-Qing and Noh, Tae Won and Lee, Daesu},
  journal={Nature Communications},
  volume={11},
  number={1},
  pages={2586},
  year={2020},
  publisher={Nature Publishing Group}
}

@article{park2021flexoelectric,
  title={Flexoelectric control of physical properties by atomic force microscopy},
  author={Park, Sung Min and Wang, Bo and Chen, Long-Qing and Noh, Tae Won and Yang, Sang Mo and Lee, Daesu},
  journal={Applied Physics Reviews},
  volume={8},
  number={4},
  pages={041327},
  year={2021},
  publisher={AIP Publishing}
}

@article{gao2018atomic,
  title={Atomic-{Scale} {Measurement} of {Flexoelectric} {Polarization} at {SrTiO}$_3$ {Dislocations}},
  author={Gao, Peng and Yang, Shuzhen and Ishikawa, Ryo and Li, Ning and Feng, Bin and Kumamoto, Akihito and Shibata, Naoya and Yu, Pu and Ikuhara, Yuichi},
  journal={Physical Review Letters},
  volume={120},
  number={26},
  pages={267601},
  year={2018},
  publisher={APS}
}

@article{young2012first,
  title={First Principles Calculation of the Shift Current Photovoltaic Effect in Ferroelectrics},
  author={Young, Steve M and Rappe, Andrew M},
  journal={Physical Review Letters},
  volume={109},
  number={11},
  pages={116601},
  year={2012},
  publisher={APS}
}

@article{rodin2014strain,
  title={Strain-Induced Gap Modification in Black Phosphorus},
  author={Rodin, A. S. and Carvalho, A. and Castro Neto, A. H.},
  journal={Physical Review Letters},
  volume={112},
  pages={176801},
  year={2014},
  doi={10.1103/PhysRevLett.112.176801}
}

@article{wei2014superior,
  title={Superior mechanical flexibility of phosphorene and few-layer black phosphorus},
  author={Wei, Qun and Peng, Xihong},
  journal={Applied Physics Letters},
  volume={104},
  pages={251915},
  year={2014},
  doi={10.1063/1.4885215}
}

\clearpage

\makeatletter
\onecolumngrid@push
\begin{center}
    \large\bfseries End Matter
\end{center}
\vspace{0.5em}
\onecolumngrid@pop
\makeatother
\appendix
\setcounter{secnumdepth}{1}

\section{BPVE conductivity and finite-wave-vector behavior}\label{app:bpve}

To obtain the flexo-BPVE coefficient, we first calculate $\sigma^{\lambda \mu \nu}[\omega; \bq, \bU]$ for a number of $\bq$ and $\bU$ according to~\cite{xu2021pure}
\begingroup\small
\begin{equation}\label{eq:bpv}
\begin{aligned}
  & \sigma^{\lambda \mu \nu}[\omega; \bq, \bU] \\ = & \frac{e^3}{ \omega^2}  \int \frac{\mathrm{d} \bk}{(2\pi)^3}  \sum_{m,n,l}  \frac{f_{lm} v^{\mu}_{lm}}{\epsilon_{ml}-\hbar \omega +i\zeta} 
 \left( \frac{v^{\lambda}_{mn} v^{\nu}_{nl}}{\epsilon_{mn}+i\zeta} - \frac{v^{\nu}_{mn} v^{\lambda}_{nl} }{\epsilon_{nl}+i\zeta}
  \right)
\end{aligned}
\end{equation}
\endgroup
Here $e$ is the electron charge, $f_{mn} \equiv f_{m} - f_{n}$ and $\epsilon_{mn} \equiv \epsilon_m - \epsilon_n$ are the difference of occupation number and band energy between bands $m$ and $n$, respectively. Particularly, one should evaluate the velocity operators by using $v^{\lambda}_{mn} = \langle \Phi_{m\bk} \vert v^{\lambda} \vert \Phi_{n\bk} \rangle$, where the perturbed wave functions $\vert \Phi_{n\bk}\rangle \equiv \vert\phi_{n\bk}\rangle + \vert\Delta\phi_{n\bk}\rangle$ is dependent on $\bq$ and $\bU$.

\section{Susceptibility of second harmonics generation}\label{app:shg}

The second-order susceptibility is calculated within the length gauge~\cite{sipe2000second,sharma2003linear}. The explicit expression is given by
\begin{equation}\label{eq:shg_formula}
\begin{aligned}
	\chi^{\lambda \mu \nu}(2\omega;\omega,\omega) & = \chi_{\rm I}^{\lambda \mu \nu}(2\omega;\omega,\omega) \\ & + \chi_{\rm II}^{\lambda \mu \nu}(2\omega;\omega,\omega) + \chi_{\rm III}^{\lambda \mu \nu} (2\omega;\omega,\omega)	
\end{aligned}
\end{equation}
where
\begin{widetext}
\begin{equation}\label{eq:shg_formula_detailed}
\begin{aligned}
	\chi_{\rm I}^{\lambda \mu \nu} (2\omega;\omega,\omega) & = \frac{2e^3}{\varepsilon_0}  \int \frac{\mathrm{d}^3 \bk}{(2\pi)^3} \sum_{mnl} \frac{r_{nm}^{\lambda} r_{ml}^{\mu} r_{ln}^{\nu}}{\epsilon_{ln}-\epsilon_{ml}} \left(\frac{f_{ml}}{\epsilon_{ml}-\hbar \omega} + \frac{f_{ln}}{\epsilon_{ln} - \hbar \omega} + \frac{2f_{nm}}{\epsilon_{mn} - 2\hbar \omega} \right)	\\
\chi_{\rm II}^{\lambda \mu \nu} (2\omega;\omega,\omega) & = \frac{2e^3}{\varepsilon_0}  \int \frac{\mathrm{d}^3 \bk}{(2\pi)^3}  \left\{\sum_{mnl} \epsilon_{mn} r_{nm}^{\lambda} \{r_{ml}^{\mu} r_{ln}^{\nu} \} \left[ \frac{f_{nl}}{\epsilon_{ln}^2 (\epsilon_{ln} - \hbar \omega)} + \frac{f_{lm}}{\epsilon_{ml}^2 (\epsilon_{ml} - \hbar \omega)}\right] \right. \\
& \left. - 8i  \sum_{nm} \frac{f_{nm} r_{nm}^{\lambda} \{\hbar \Delta_{mn}^{\mu} r_{mn}^{\nu}\}}{\epsilon_{mn}^2 (\epsilon_{mn} - 2\hbar \omega) }  
 - 2 \sum_{nml} \frac{f_{nm} r_{nm}^{\lambda}  \{r_{ml}^{\mu} r_{ln}^{\nu} \}(\hbar\omega_{ln}-\hbar\omega_{ml})}{\epsilon_{mn}^2 (\epsilon_{mn} - 2\hbar \omega)} \right\} \\
\chi_{\rm III}^{\lambda \mu \nu} (2\omega;\omega,\omega) & = \frac{e^3}{\varepsilon_0}  \int \frac{\mathrm{d}^3 \bk}{(2\pi)^3}   \left\{ \sum_{nml} \frac{f_{nm}}{\epsilon_{mn}^2 (\epsilon_{mn} - \hbar \omega)} \left[ \epsilon_{nl} r_{lm}^{\lambda} \{r_{mn}^{\mu} r_{nl}^{\nu}\} - \epsilon_{lm} r_{nl}^{\lambda} \{r_{lm}^{\mu} r_{mn}^{\nu} \} \right] \right. \\
& + \left. i \sum_{nm} \frac{f_{nm} r_{nm}^{\lambda}}{\epsilon_{mn}^2 (\epsilon_{mn} - \hbar \omega)} \{\hbar{\Delta}_{mn}^{\mu} r_{mn}^{\nu}\} \right\} 
\end{aligned}
\end{equation}
\end{widetext}
Here the inter-band position matrix is defined as $r_{mn}^{\lambda} = \frac{\hbar v_{mn}^{\lambda}}{i \epsilon_{mn}}$ for $m\neq n$ and $r_{mn}^{\lambda}=0$ for $m=n$. $\Delta_{mn}^{\lambda} \equiv v_{mm}^{\lambda} - v_{nn}^{\lambda}$ is the difference in band velocities. Meanwhile, for two numbers $A$ and $B$, one has $\{AB\} \equiv \frac{1}{2}(AB+BA)$. Other quantities have been defined in the main text.

\end{document}